\documentclass[floatfix,reprint,footinbib,twocolumn,prb,a4paper,preprintnumbers,amsmath,amssymb]{revtex4-1}

\usepackage{graphics}
\DeclareGraphicsExtensions{.png,.pdf}
\usepackage{epsfig}
\usepackage{color}
\usepackage{hyperref}
\usepackage[utf8]{inputenc}

\begin{document}
\title{\bf First-order ferromagnetic transitions of lanthanide local moments in divalent compounds: An itinerant electron positive feedback mechanism and Fermi surface topological change}
\author{Eduardo Mendive-Tapia,$^{1,2}$ Durga Paudyal,$^{3}$ Leon Petit,$^{4}$ and Julie B.~Staunton,$^{2}$}
\affiliation{$^{1}$Department of Computational Materials Design, Max-Planck-Institut für Eisenforschung GmbH, 40237 Düsseldorf, Germany}
\affiliation{$^{2}$Department of Physics, University of Warwick, Coventry CV4 7AL, United Kingdom}
\affiliation{$^{3}$The Ames Laboratory, U.~S.~Department of Energy, Iowa State University, Iowa 50011-1015, USA}
\affiliation{$^{4}$Daresbury Laboratory, Warrington WA4 4AD, United Kingdom}
\date{\today}

\begin{abstract}

Around discontinuous (first-order) magnetic phase transitions the strong caloric response of materials to the application of small fields is widely studied for the development of solid-state refrigeration. Typically strong magnetostructural coupling drives such transitions and the attendant substantial hysteresis dramatically reduces the cooling performance. In this context we describe a purely electronic mechanism which pilots a first-order paramagnetic-ferromagnetic transition in divalent lanthanide compounds and which explains the giant non-hysteretic magnetocaloric effect recently discovered in a Eu$_2$In compound. There is positive feedback between the magnetism of itinerant valence electrons and the ferromagnetic ordering of local $f$-electron moments, which appears as a topological change to the Fermi surface. The origin of this electronic mechanism stems directly from Eu’s divalency, which explains the absence of a similar discontinuous transition in Gd$_2$In.

\end{abstract}

\maketitle

\section{Introduction}
\label{intro}

Local moments formed on atoms from strongly-correlated $f$-electrons interact with each other to produce a plethora of magnetic phases in lanthanide compounds~\cite{Mackintosh1}. They typically interact by spin-polarizing the itinerant pervasive valence electron sub-system in line with the famous Ruderman-Kittel-Kasuya-Yoshida (RKKY) paradigm~\cite{Mackintosh1,Hughes1}. The transitions between magnetic phases are frequently continuous (second-order). In many cases, however, the magnetic state is strongly coupled to the crystal structure and discontinuous (first-order) magnetic phase transitions occur along with sudden crystal volume~\cite{Lewis_2016,doi:10.1063/1.1375836,PhysRevB.67.104416} and symmetry~\cite{PhysRevLett.78.4494,PhysRev.126.104} change.
In principal, magnetic materials which undergo such discontinuous effects are very valuable for technological applications - 
 response to small or moderate external fields can lead to large changes of temperature, volume, magnetization, and other thermodynamic quantities. Such functionality can exploit strong magnetoresistance~\cite{doi:10.1063/1.2399365}, magnetostriction~\cite{Nikitin_2011}, magnetic shape memory~\cite{ShapeMemoKainuma,ShapeMemoManosa}, and caloric effects for solid-state cooling~\cite{PhysRevLett.78.4494,ShapeMemoManosa,Matsunami1,PhysRevX.8.041035}.

\begin{figure}[t]
\centering
\includegraphics[clip,scale=0.77]{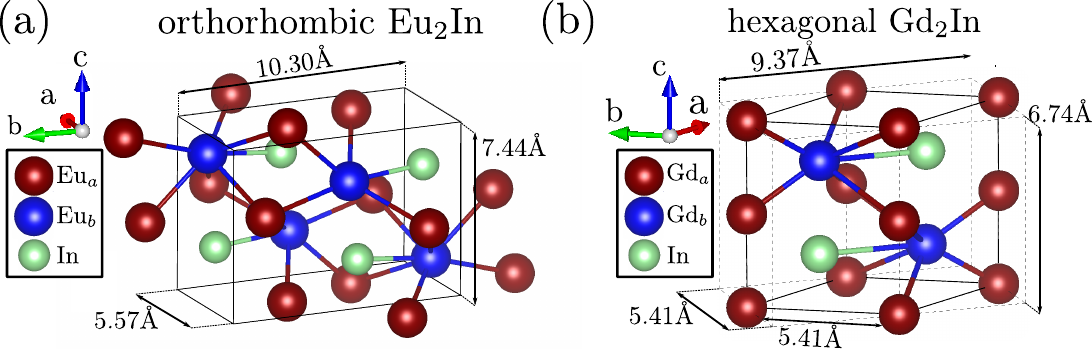}
\caption{Crystallographic unit cells of (a) orthorhombic Eu$_2$In and (b) hexagonal Gd$_2$In systems. Non-equivalent RE positions, RE$_a$ and RE$_b$, are drawn using red and blue, respectively.}
\label{FIG1}
\end{figure}

First-order magnetic phase transitions, however, have a downside in that they are almost always accompanied by significant hysteresis~\cite{doi:10.1002/ente.201800264}.
A striking exemption is the Eu$_2$In compound~\cite{Eu2In1} which exhibits a phase transition from a paramagnetic (PM) to a ferromagnetic (FM) phase at a Curie temperature of $T_c\approx$ 55K. Although the transformation is strongly first-order with an associated giant magnetocaloric effect (MCE), it exceptionally shows no thermal hysteresis~\cite{Eu2In1}.
The transition is isosymmetric for this orthorhombic crystalline material (Fig.\ \ref{FIG1}(a)) and no substantial change of lattice parameters occurs at the magnetic discontinuity. This implies negligible magnetostructural coupling and a possible purely electronic origin for the transformation. 
PM-FM magnetic phase transitions observed in other RE$_2$In compounds (RE -rare earth)~\cite{McAlister_1984,PALENZONA1968379} are all second-order with smaller, although still significant, MCE~\cite{doi:10.1063/1.362319,BHATTACHARYYA20121239,ZHANG2009396,BHATTACHARYYA20091828,Zhang_2009,ANH2006132,refId0,doi:10.1063/1.3130090}. With the exceptions of Eu and Yb, RE$_2$In compounds generally crystallize into a hexagonal Ni$_2$In-type structure (Fig.\ \ref{FIG1}(b)), which is a continuous distortion of the orthorhombic structure observed in Eu$_2$In. 

Lanthanide atoms are typically trivalent in a solid (nominal 5$d^1$6$s^2$ valence electron configuration), donating three valence electrons per atom to the electron glue in which the atomically-localized $f$-electron magnetic moments sit~\cite{Mackintosh1}. The gain in $f$-electron correlation energy that results from a half-filled $f$-electron shell makes an exception of Eu, steering it towards divalency instead and producing a magnetic moment per atom of $\approx$ 7 $\mu_B$ from the seven localized $f$-electrons~\cite{Eu2In1}. In this paper we show with an {\it ab initio} theory that the divalency causes the Fermi energy $E_\text{F}$ of Eu$_2$In to be positioned so that there is a positive feedback from the itinerant valence electron sub-system on the magnetic interactions among the $f$-electron magnetic moments. This results in a first-order PM-FM transition which is devoid of any magnetostructural coupling. 

The electronic mechanism we have discovered appears in both orthorhombic and hypothetical hexagonal Eu$_2$In but is absent in any of the other RE$_2$In counterparts such as Gd$_2$In.  We have found that the itinerant electron positive feedback driving the first-order character in Eu$_2$In is caused by a Fermi surface (FS) topological transition. Around $E_\text{F}$ there are Eu $d$ - In $p$ electron bonding states with pronounced $d$-character and strong susceptibility to spin polarization. This illustrates a general point that in a material where a rare earth metal can be be coaxed into a divalent state its valence d-electrons will participate in a rich range of magnetic interactions.  In Eu$_2$In the change of FS topology is brought about by the spin-polarization of these states produced by the magnetic field which is set up by the alignment of Eu-moments as ferromagnetic order develops. In turn it further strengthens the interactions among the moments, underlying the origin of the first-order transformation.

The paper is organized as follows. In section \ref{mechanism} we describe the purely electronic mechanism driving the discontinuous magnetic phase transition of Eu$_2$In using a first-principles disordered local moment theory. These results are compared with those for trivalent Gd$_2$In in order to illustrate how the divalent state of the rare earth underpins the first-order character. We describe the role of the FS topological change in subsection \ref{FS}. Section \ref{free} shows how the magnetic phase transitions are described in the theory along with temperature-dependent magnetocaloric properties via the minimization of a first-principles Gibbs free energy. Further details on the theoretical formalism are given in section \ref{formalism}. Finally, in section \ref{conc} we present our conclusions.

\section{Itinerant electron positive feedback mechanism}
\label{mechanism}

\begin{figure}[t]
\centering
\includegraphics[clip,scale=0.68]{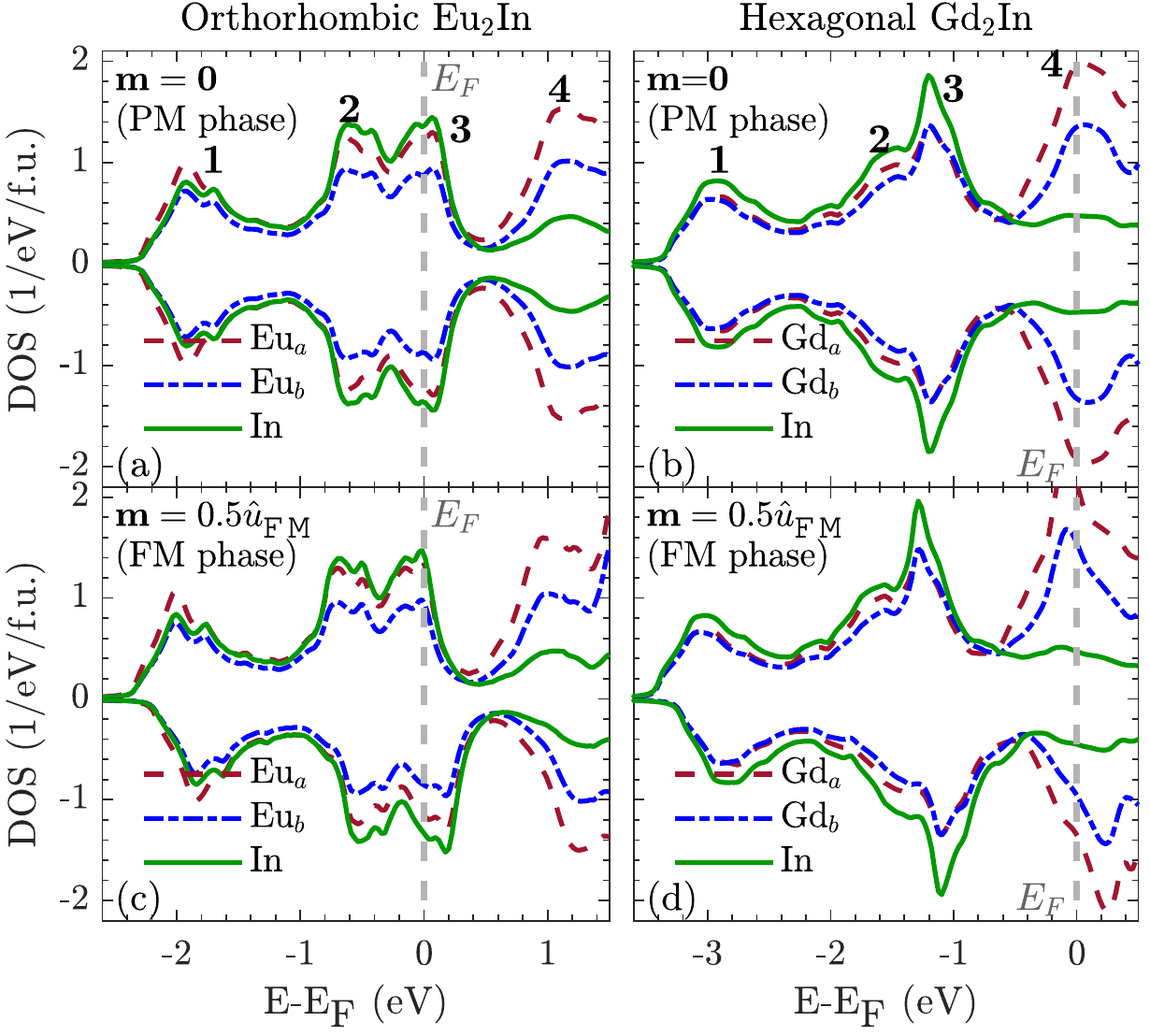}
\caption{Sub-lattice-resolved spin-polarized densities of states (DOS) of the $s$, $p$ and $d$ valence electrons of orthorhombic Eu$_2$In (left panels) and hexagonal Gd$_2$In (right panels). RE non-equivalent sub-lattices are labeled $a$ and $b$. The upper panels show the DOS for the paramagnetic DLM state where the magnetic order parameters $\textbf{m}_a=\textbf{m}_b=0$. The lower panels show the DOS for partially ferromagnetically ordered compounds with $\frac{\textbf{m}_a+\textbf{m}_b}{2}=0.5\hat{u}_\text{FM}$. In each figure positive (negative) values correspond to the average of the DOS of electrons spin-polarized parallel (anti-parallel) to the direction of overall magnetization of the system,  $\hat{u}_\text{FM}$. Some features in the DOSs are shown by numerical labels. The Fermi energy $E_\text{F}$ is indicated by a vertical gray dashed line.}
\label{FIG2}
\end{figure}

The evidence for the itinerant electron positive feedback mechanism starts with the upper panels of Fig.\ \ref{FIG2}(a,b). The figure shows the sub-lattice resolved and spin polarized densities of states (DOS) of the valence electrons above $T_c$ in the PM state for both orthorhombic Eu$_2$In and hexagonal Gd$_2$In, obtained using our density functional theory (DFT)-based disordered local moment (DLM) theory~\cite{0305-4608-15-6-018,Hughes1}. We have modelled the local $f$-electron moments as being randomly orientated (in a disordered local moment, DLM, state) so that there is no overall magnetization. The FM phase at lower temperatures, shown in Fig.\ \ref{FIG2}(c,d), is described by the magnetic order parameter of the system, $\textbf{m} =\frac{\textbf{m}_a+\textbf{m}_b}{2}= \frac{\langle \hat{e}_a \rangle + \langle \hat{e}_b \rangle}{2}$, where $\hat{e}_{a(b)}$ denotes an orientation of a local moment on a site on the RE sublattice $a (b)$, $\textbf{m}_{a(b)}\equiv\langle\hat{e}_{a(b)}\rangle$ is a local order parameter, and so $\textbf{m}$ specifies the corresponding total magnetization average. Note that for the PM state $\textbf{m}_{a(b)}=\textbf{0}$, i.e.\ $\textbf{m}=\textbf{0}$, and for the full FM order $\textbf{m}=1 \hat{u}_\text{FM}$, where $\hat{u}_\text{FM}$ is a unit vector along the overall magnetization direction of the system at zero-temperature. For both the divalent Eu and trivalent Gd atoms we have obtained local moments which are roughly 7$\mu_B$ in magnitude consistent with their half filled atomic $f$-shells ($\mu_{\text{Eu}_a}=7.17\mu_\text{B}$ and $\mu_{\text{Eu}_b}=7.12\mu_\text{B}$ for Eu$_2$In, and $\mu_{\text{Gd}_a}=7.23\mu_\text{B}$ and $\mu_{\text{Gd}_b}=7.13\mu_\text{B}$ for Gd$_2$In). The positive (negative) values in the figure show $+$ ($-$) the average of the DOS spin-polarized parallel (anti-parallel) to an overall magnetization direction $\hat{u}_\text{FM}$. In the PM states ($\textbf{m}=\textbf{0}$) the majority and minority spin DOS are identical reflecting the zero overall spin polarization of their electronic structures.  In the ferromagnetically-ordered states ($\textbf{m}\neq\textbf{0}$) the $f$-electron local moments have orientational bias towards $\hat{u}_\text{FM}$ which produces an internal magnetic field. As a consequence the electronic structures averaged over the local moment orientations show spin-polarization (Fig.\ \ref{FIG2}(c,d)). 

\begin{figure*}[t]
\centering
\includegraphics[clip,scale=0.27]{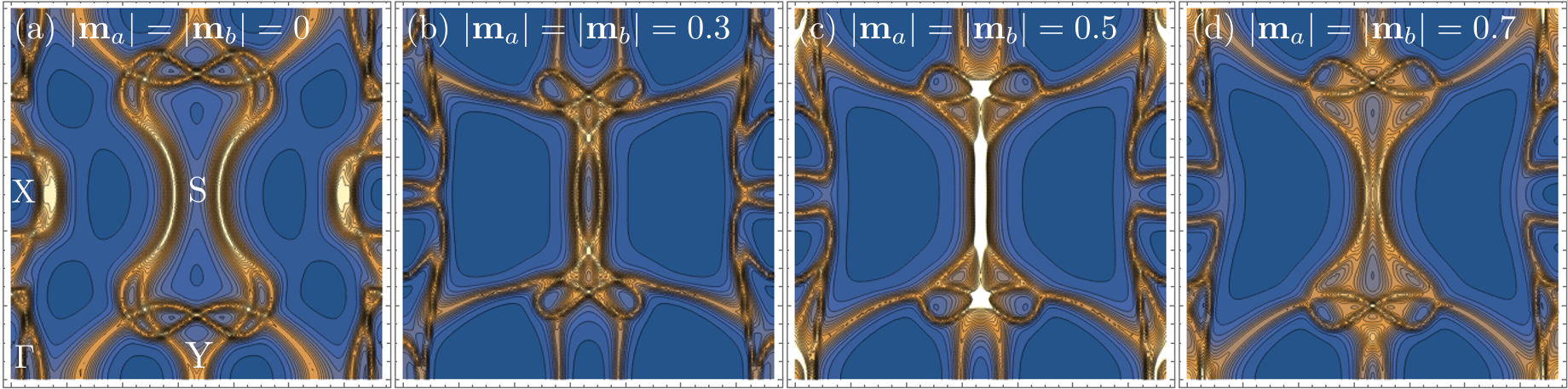}
\caption{
(a) Eu$_2$In's Bloch spectral function $A_B (\textbf{k}= (k_x,k_y,0),E_\text{F})$ at the Fermi energy for the majority spin electrons in the (a) PM ($|\textbf{m}_a|=|\textbf{m}_b|=0$) and (b-d) FM states for values of the magnetic order parameters $|\textbf{m}_a|=|\textbf{m}_b|=0.3$, $0.5$ (near the first-order PM-FM transition (see Figs.\ \ref{FIG5} and \ref{FIG6})) and $0.7$, respectively. Panel (a) shows characteristic reciprocal space points ($\Gamma$, S, X, Y). $k_x$ ($k_y$) is the value along the $x$ axis ($y$ axis) in the figure and specifies the $b$ axis ($a$ axis) in the crystal.}
\label{FIG3}
\end{figure*}

\subsection{Fermi surface topological change}
\label{FS}

The electronic structures of the two materials have similar general features which can be gauged by the labels 1, 2, 3 and 4 in Fig.~\ref{FIG2}. There are RE-In bonding states incorporating RE $d$ and In $p$ electron hybridization at the lower energies (labels 2-3). Non-bonding unhybridized RE and In states are observed at higher energies (label 4). The low energy RE $d$ - In $p$ band's susceptibility to being spin polarized by the internal magnetic field, which is set up when the lanthanide $f$-moments align in the FM state, can be seen from the nearly rigid shift of features of the majority and minority spin electron DOS.  The spin splitting along the energy axis of the DOS for a partially ordered FM state ($\textbf{m}=0.5\hat{u}_\text{FM}$) is approximately 0.1eV. The game changing difference between the two materials is the lower position of Eu$_2$In's  Fermi energy compared to that of Gd$_2$In simply because Eu is divalent whereas Gd is trivalent. Eu$_2$In's  Fermi energy lies near the top of the Eu-In $pd$ bonding complex such that the interaction between the Eu local $f$-moments is mediated by these electrons. When the system ferromagnetically orders the majority spin DOS fills further while the minority spin states depopulate.  On the other hand, $E_\text{F}$ of Gd$_2$In is much higher, well out of this $pd$ structure.

The interactions between RE local moments in many intermetallics depend on the response of the surrounding valence electrons to the local magnetic fields produced by the moments. Such an itinerant electron spin susceptibility is proportional to the DOS at the Fermi energy, $n(E_\text{F})$. In Eu$_2$In the states around $E_\text{F}$ are rigidly exchange-split by the magnetic field set up by the overall FM order of the RE moments and so $n(E_\text{F})$ develops a marked dependence on $\textbf{m}$, if $E_\text{F}$ is near the top or bottom of a band. There are features analogous with itinerant electron metamagnetism as described by Wohlfarth and Rhodes~\cite{doi:10.1080/14786436208213848} and reported by Fujita \textit{et al}.\ as a factor behind the first-order PM-FM transition in La(Fe$_x$Si$_{1-x}$)$_{13}$ materials, which shows small hysteresis, and other compounds~\cite{doi:10.1063/1.370471,OHTA2005431}. A similar effect causes Eu$_2$In's DOS at $E_\text{F}$ to increase as the overall magnetization $\textbf{m}$ grows. This ultimately strengthens the magnetic interactions and drives the first-order PM-FM transition. In Eu$_2$In this is the only relevant effect with a negligible coupling to the crystal structure and a full removal of hysteresis.

\begin{figure}[t]
\centering
\includegraphics[clip,scale=0.35]{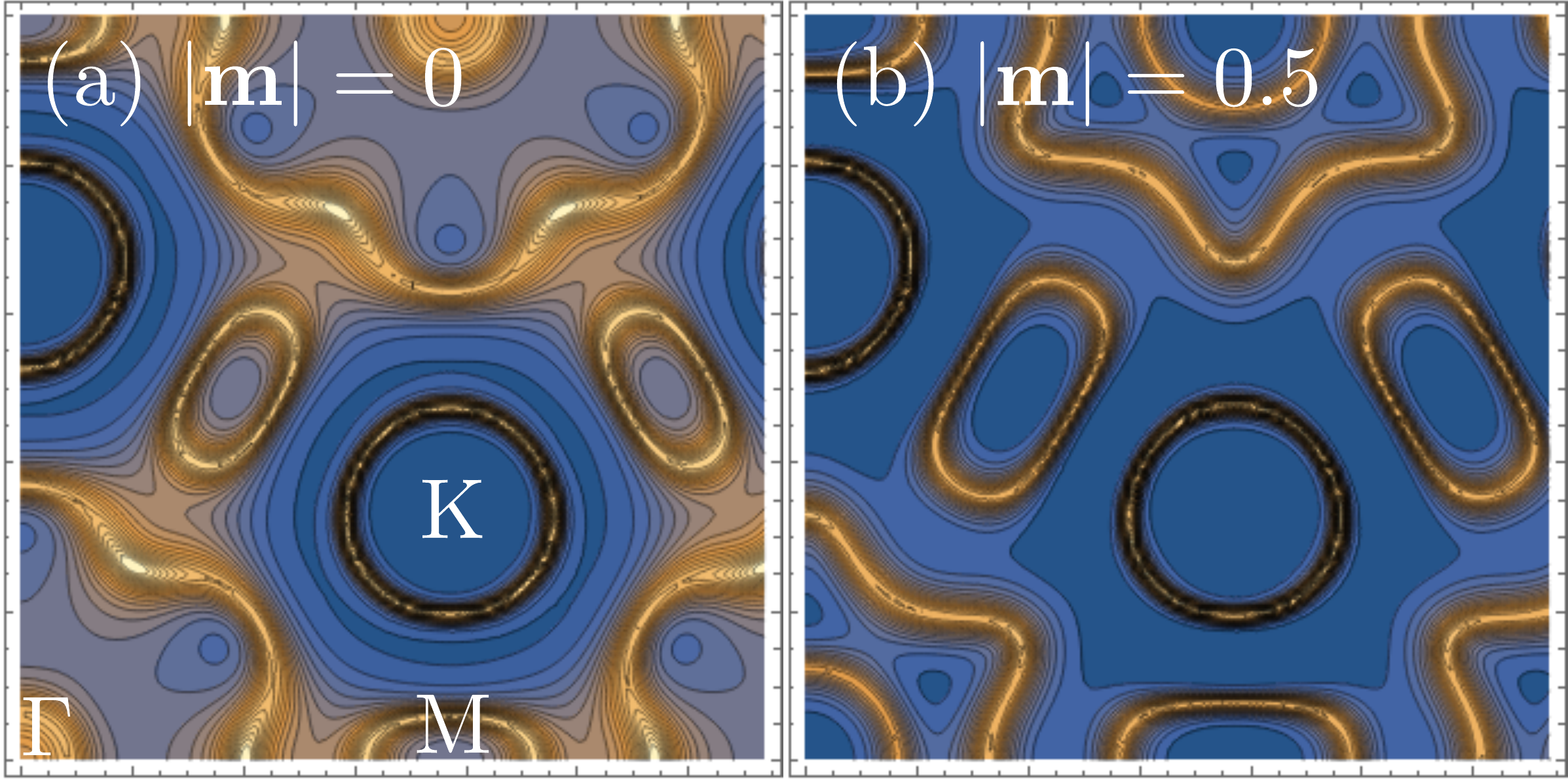}
\caption{
(a) Gd$_2$In's Bloch spectral function $A_B (\textbf{k}= (k_x,k_y,0),E_\text{F})$ at the Fermi energy for the majority spin electrons in the (a) PM ($|\textbf{m}_a|=|\textbf{m}_b|=0$) and (b) a partially ordered ferromagnetic state ($|\textbf{m}|=0.5$). Panel (a) shows characteristic reciprocal space points ($\Gamma$, M, K).}
\label{FIG4}
\end{figure}

Further scrutiny of the Fermi surface uncovers how this itinerant electron positive feedback effect occurs. Examination of the wave-vector $\textbf{k}$-dependence of the Bloch spectral function, $A_B (\textbf{k}= (k_x,k_y,k_z),E_\text{F})$,~\cite{PhysRevB.21.3222,PhysRevB.73.205109} reveals van Hove singularities in the  majority spin electronic structure of Eu$_2$In when the $f$ local moment FM order becomes large enough (see Fig.\ \ref{FIG3}(c)). Eu $d$ - In $p$ hybridized states become sufficiently spin-polarized to cause a topological change to the majority spin FS. This occurs when the Eu local moment magnetic order exceeds a critical value and a valence band fills (see Fig.\ \ref{FIG2}). Fig.~\ref{FIG3} illustrates the effect by a comparison of the Fermi surface of the valence electrons in the PM state ($\textbf{m}=\textbf{0}$) with the majority spin Fermi surface when there is increasing partial ferromagnetic order of local Eu $f$-moments, i.e.\ for increasing values of $\textbf{m}$. For the $k_z=0$ plane of the PM Fermi surface (Fig.\ref{FIG3}(a)), there is a dog bone structure aligned with the $a$ axis with its middle at $k_x= \frac{\pi}{a}$. With sufficient $f$-moment FM order ($\textbf{m}=0.5 \hat{u}_\text{FM}$) some majority spin $pd$ bands become completely occupied so that an  intense feature in the Bloch spectral function manifests at the Brillouin zone boundary perpendicular to the $b$ axis (Fig.\ref{FIG3}(c)). Substantial Fermi surface nesting (Kohn anomalies) occurs for reciprocal lattice vectors along the $b$ axis which enhances the magnetic interactions stabilizing ferromagnetism further. 
On the other hand, Gd$_2$In's Fermi surface shows more free electron-like features compliant with its more conventional RKKY-like interactions. Fig.\ \ref{FIG4} shows that in Gd$_2$In such a topological change is absent and only a small smearing occurs as ferromagnetic order varies.

\section{First Principles Theory for the Gibbs free energy}
\label{free}

The {\it ab initio} theory used here subsumes the valence electron effects presented in section \ref{FS}. Both Eu$_2$In and Gd$_2$In have the relative simplicity of an S-state for each  lanthanide atom's localized $f$-electrons so that crystal field and spin-orbit coupling effects are small. The relevant degrees of freedom are described, taking the large S limit, as unit vectors $\{ \hat{e}_n\}$ specifying the orientations of the local $f$-electron spin moments at crystal sites $\{n\}$, with averages $\{\textbf{m}_n=\langle\hat{e}_n\rangle\}$. These moment orientations fluctuate on a time scale much longer than other electronic motions and each configuration $\{\hat{e}_n\}$ imposes a transient non-collinear spin-polarization on the systems' valence electrons. As shown elsewhere~\cite{PhysRevB.99.144424,PhysRevB.89.054427} from this premise an \textit{ab initio} Gibbs free energy can be constructed, 
\begin{equation}
\mathcal{G}_1= \Omega \big(\{\textbf{m}_n\},\textbf{H},T \big)-TS,
\label{EQ1}
\end{equation}
where $\Omega\big(\{\textbf{m}_n\},\textbf{H},T\big)$ is the magnetic energy of the material, which can include the effect of an external magnetic field $\textbf{H}$, and $S=\sum_n S_n$ is the total entropy of the local moments.
$\Omega \big(\{\textbf{m}_n\},\textbf{H},T\big)$ is obtained as an average over local moment configurations of the grand potential, $\Omega_c(\{\hat{e}_n\},\textbf{H},T)$, of the interacting electrons with spin polarization constrained to $\{\hat{e}_n\}$~\cite{0305-4608-15-6-018}. The equilibrium state of the system for specific values of the temperature, $T$, and $\textbf{H}$, is given by the set of order parameters $\{\textbf{m}_n\}$ which minimizes the Gibbs free energy function $\mathcal{G}_1$~\cite{PhysRevB.99.144424}. See subsection \ref{formalism} for further details on the first-principles calculation of $\mathcal{G}_1$.

The separation of electronic degrees of freedom into relatively slow local moments and the faster remaining ones introduces two distinct temperature effects. There is the thermal disordering of the local moment orientations described by an explicit dependence on $\{\textbf{m}_n\}$, and also the particle-hole excitations within each spin polarization-constrained many electron system. Formally, the grand potential $\Omega_c(\{\hat{e}_n\},\textbf{H},T)$ is provided by density functional theory~\cite{PhysRev.137.A1441,0305-4608-15-6-018}. It is a functional of the electronic charge and magnetization densities and minimizes with respect to them subject to the spin polarization constraint $\{\hat{e}_n\}$~\cite{0305-4608-15-6-018}.  The sums over Kohn-Sham single electron energies to produce the charge and magnetization densities and other quantities are weighted by the Fermi-Dirac distribution and depend, therefore, on $T$. Since $\Omega \big(\{\textbf{m}_n\},\textbf{H},T \big)$ is the average of $\Omega_c(\{\hat{e}_n\},\textbf{H},T)$ over local moment configurations, it includes a temperature dependence from single electron-hole excitations. This dependence filters through the calculation of magnetic phase diagrams and can introduce a sizable contribution to caloric properties~\cite{PhysRevB.89.054427}. It also makes the interactions between the local moments temperature-dependent in principle as we find for Eu$_2$In.  

\begin{figure}[t]
\centering
\includegraphics[clip,scale=1]{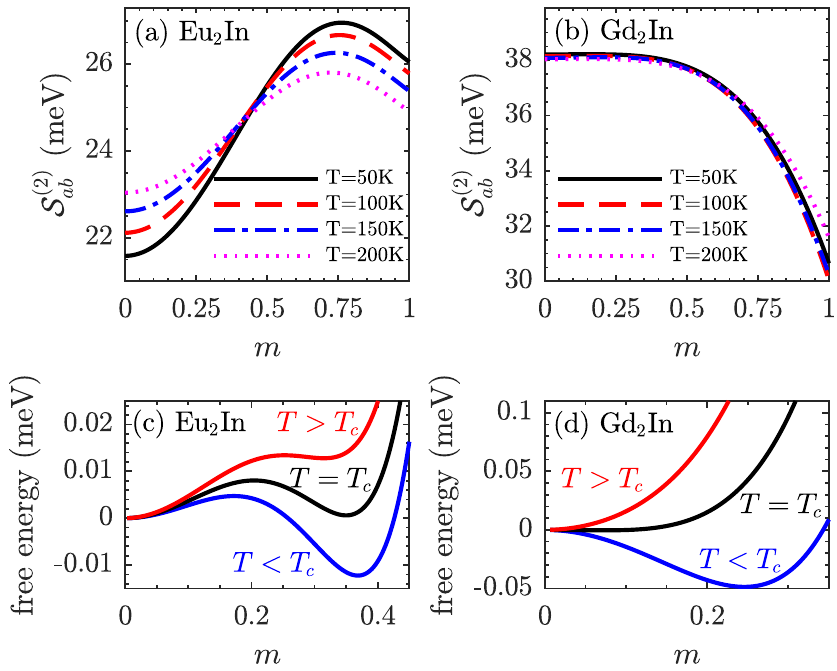}
\caption{(a,b) Dominant magnetic interactions, $\mathcal{S}^{(2)}_{ab}$, in Eu$_2$In and Gd$_2$In between sub-lattices $a$-$b$ as functions of ferromagnetic order $m=|\textbf{m}|$ and for different temperatures. (c,d) Gibbs free energies, $\mathcal{G}_1$, of these compounds against $m$ above, at, and below their Curie temperatures $T_c$.}
\label{FIG5}
\end{figure}

As demonstrated in our studies of the magnetic phases and caloric properties of FeRh alloys~\cite{PhysRevB.89.054427}, heavy lanthanide elements~\cite{PhysRevLett.118.197202} and Mn-rich materials~\cite{PhysRevB.99.144424, PhysRevB.95.184438}, the dependence of $ \mathcal{G}_1$ on $\{\textbf{m}_n\}$ is established by a linear regression analysis of the magnetic energy $\Omega$, which is described in section \ref{formalism}, calculated for a large number of prescribed averages $\{\textbf{m}_n\}$ and fixed $T$. For Eu$_2$In and Gd$_2$In, $\mathcal{G}_1$ has lowest values for ferromagnetic configurations, i.e.\ it minimizes with respect to $\{\textbf{m}_a,\textbf{m}_b\}$ describing a FM phase. The internal energy per unit cell is fit well by the expression
\begin{eqnarray}
& & \Omega \big(\textbf{m}_a,\textbf{m}_b,\textbf{H},T \big) = \Omega_0- \textbf{H} \cdot (\textbf{m}_a + \textbf{m}_b) \nonumber \\
& & -\mathcal{S}^{(2)}_{aa}(\textbf{m}_{a},\textbf{m}_{b},T)\textbf{m}_a\cdot\textbf{m}_a -\mathcal{S}^{(2)}_{bb}(\textbf{m}_{a},\textbf{m}_{b},T)\textbf{m}_b\cdot\textbf{m}_b  \nonumber \\
& & -\mathcal{S}^{(2)}_{ab}(\textbf{m}_{a},\textbf{m}_{b},T)\textbf{m}_a\cdot\textbf{m}_b,  
\label{EQ2}
\end{eqnarray}
where $\Omega_0$ is a constant, and $\mathcal{S}^{(2)}_{aa}$, $\mathcal{S}^{(2)}_{ab}$ and $\mathcal{S}^{(2)}_{bb}$ compactly comprise the magnetic interactions between sub-lattices $a\text{-}a$, $b\text{-}b$, and $a\text{-}b$, respectively. They depend on the extent of the overall ferromagnetic order of the system, $\textbf{m}=\frac{\textbf{m}_{a}+\textbf{m}_{b}}{2}$, and also explicitly on the temperature via the electron-hole excitations. We have found that for both Eu$_2$In and Gd$_2$In the intersublattice interactions $\mathcal{S}^{(2)}_{ab}$ dominate and determine the magnetic phase behavior. Their magnetic order parameter and temperature dependencies obtained in section \ref{formalism} are shown in Fig~\ref{FIG5}(a,b). Eu$_2$In's $\mathcal{S}^{(2)}_{ab}$ increase significantly with increasing $\{\textbf{m}_{a},\textbf{m}_{b}\}$  up to magnitudes of $m=\left|\frac{\textbf{m}_a+\textbf{m}_b}{2}\right|\approx 0.6$ before dropping back slightly for larger values. This behavior tracks the DOS at $E_\text{F}$ and the FS evolution with $m$ shown in Figs.~\ref{FIG2}(a,c) and \ref{FIG3} and captures the itinerant electron positive feedback in Eu$_2$In. The greatest rate of increase coincides with the change of FS topology around $\textbf{m} = 0.5\hat{u}_\text{FM}$.
Inspecting the explicit dependence of the magnetic interactions on temperature shows that lowering $T$ strengthens the FS role. This means that the dependence of the local moment interactions on the extent of the spin polarization of the valence electrons increases as $T$ decreases. For Gd$_2$In opposite trends are found in line with the absence of the mechanism for this material - its $\mathcal{S}^{(2)}_{ab}$ is insensitive to increasing $m$ up to values of 0.6 before decreasing, and shows negligible temperature dependence. 

\begin{figure}[t]
\centering
\includegraphics[clip,scale=0.78]{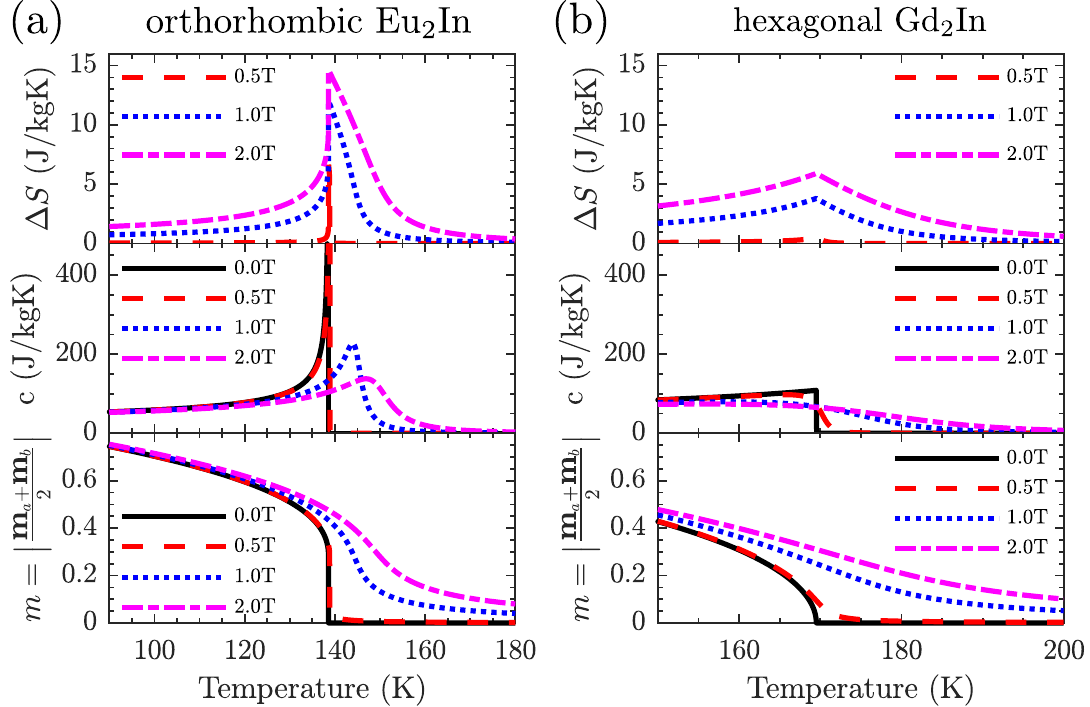}
\caption{The heat capacity, $c$, isothermal magnetic entropy change, $\Delta S$, and total magnetic order parameter, $|\textbf{m}|=m$, in applied magnetic fields up to 2 Tesla calculated from the theory for (a) Eu$_2$In and (b) Gd$_2$In. The sharp PM-FM first-order transition found in experiment in Eu$_2$In is evident, although at a somewhat higher temperature (140K rather than 55K). The same theory finds a second-order transition in Gd$_2$In in agreement with experiment.}
\label{FIG6}
\end{figure}

In Fig.~\ref{FIG5}(c,d) the free energy $\mathcal{G}_1$ is plotted against $m$, the extent of ferromagnetic order.  Eu$_2$In's plot shows a clear discontinuous PM-FM phase transition at $T_c =140$K. This leads to the first-order behavior of the heat capacity and magnetization which we plot in Fig.\ \ref{FIG6}(a). Fig.\ \ref{FIG6}(a) also shows its corresponding MCE, described by the entropy change $\Delta S$ produced by the change of magnetic order and obtained by minimizing $\mathcal{G}_1$. $\Delta S$ is giant and its value is in good qualitative agreement with experimental measurements at 1 and 2 Tesla~\cite{Eu2In1}. 
Eu$_2$In's free energy, magnetization, heat capacity and MCE are in sharp contrast with the second-order behavior obtained for Gd$_2$In and with its smaller MCE, as depicted in Figs.~\ref{FIG5}(d) and \ref{FIG6}(b).
Although the Curie transition temperatures $T_c$ provided by our first-principles theory are above experimental findings, their values are low and capture the experimental trend of $T_c\text{(for Eu$_2$In)}<T_c\text{(for Gd$_2$In)}$. We point out that our treatment of the overall effect of the magnetic interactions is based on a mean-field theory, which typically overestimates the value of $T_c$.

\subsection{Details of Theoretical Formalism and Calculational Method}
\label{formalism}

The calculation of the first-principles internal magnetic energy in Eq.\ (\ref{EQ2}), $\Omega\big(\textbf{m}_a,\textbf{m}_b,\textbf{H},T \big)$, is the central component of the Gibbs free energy presented in Eq.\ (\ref{EQ1}) above. We calculate $\Omega\big(\textbf{m}_a,\textbf{m}_b,\textbf{H},T \big)$ by constructing a mean-field theory to efficiently describe different magnetically constrained states of the grand potential $\Omega_c(\{\hat{e}_n\},\textbf{H},T)$, which provides the average $\Omega=\langle\Omega_c(\{\hat{e}_n\},\textbf{H},T)\rangle$.
A simpler trial Hamiltonian $\mathcal{H}_0=-\sum_n\textbf{h}^\text{int}_n\cdot\hat{e}_n$ setting a site-dependent magnetic field $\textbf{h}^\text{int}_n$ is used to capture the overall effect of the magnetic interactions~\cite{0305-4608-15-6-018}.
An advantage of our approach is that averages over the phase space of magnetic configurations $\{\hat{e}_n\}$ with respect to the corresponding single-site trial probability distribution, $\{P_n(\hat{e}_n)=\exp\left[\beta\textbf{h}^\text{int}_n\cdot\hat{e}_n\right]\}$ where $\beta=1/k_\text{B}T$ is the Boltzmann factor, can be performed using the Coherent Potential Approximation (CPA)~\cite{PhysRev.156.809,PhysRevB.5.2382}. This technology is implemented within the Multiple Scattering Theory (MST) formalism of DFT known as the Korringa-Kohn-Rostoker (KKR) method~\cite{KORRINGA1947392,PhysRev.94.1111,doi:10.1080/00018737200101268}.
The probability of each configuration is given by $P(\{\hat{e}_n\}) = \prod_nP_n(\hat{e}_n)$, and the corresponding partition function is $Z_0 =\prod_n{Z_{0,n}}=\prod_n{\int{\text{d}\hat{e}_n\exp\left[\beta\textbf{h}^\text{int}_n\cdot\hat{e}_n\right]}}=\prod_n{4\pi\frac{\sinh\beta h^\text{int}_n}{\beta h^\text{int}_n}}$~\cite{0305-4608-15-6-018}. The local moment order parameters, which are the inputs in our theory, are therefore given by
\begin{equation}
\bigg\{\textbf{m}_n\equiv\langle\hat{e}_n\rangle=\int\text{d}\hat{e}_n P_n(\hat{e}_n)\hat{e}_n
=L(\beta h^\text{int}_n)\hat{h}^\text{int}_n\bigg\},
\label{EQmn}
\end{equation}
where $L(\beta h^\text{int}_n)=\frac{-1}{\beta h^\text{int}_n}+\coth(\beta h^\text{int}_n)$ is the Langevin function of $\beta h^\text{int}_n$.

We have found that the dependence of $\{\mathcal{S}_{aa},\mathcal{S}_{bb},\mathcal{S}_{ab}\}$ on $\{\textbf{m}_a,\textbf{m}_b\}$ in Eq.\ (\ref{EQ2}) can be described well by
\begin{equation}
\begin{split}
\mathcal{S}^{(2)}_{aa}=\mathcal{S}^{(2)}_{aa,0} & +\mathcal{S}^{(4)}_{aa}\textbf{m}_a\cdot\textbf{m}_a+\mathcal{S}^{(6)}_{aa}(\textbf{m}_a\cdot\textbf{m}_a)^2, \\
\mathcal{S}^{(2)}_{bb}=\mathcal{S}^{(2)}_{bb,0} & +\mathcal{S}^{(4)}_{bb}\textbf{m}_b\cdot\textbf{m}_b+\mathcal{S}^{(6)}_{bb}(\textbf{m}_b\cdot\textbf{m}_b)^2, \\
\mathcal{S}^{(2)}_{ab}=\mathcal{S}^{(2)}_{ab,0} & +\mathcal{S}^{(4)}_{ab,1}\textbf{m}_a\cdot\textbf{m}_a+\mathcal{S}^{(4)}_{ab,2}\textbf{m}_b\cdot\textbf{m}_b \\
 & +\mathcal{S}^{(4)}_{ab,3}\textbf{m}_a\cdot\textbf{m}_b+\mathcal{S}^{(6)}_{ab,1}(\textbf{m}_a\cdot\textbf{m}_a)^2 \\
 & +\mathcal{S}^{(8)}_{ab,1}(\textbf{m}_a\cdot\textbf{m}_a)^3
.
\label{EQSm}
\end{split}
\end{equation}
In our theory such a magnetic order dependence generates multi-site local moment correlations, i.e.\ higher order than pairwise, in the free energy $\mathcal{G}_1$, as can be seen by introducing Eq.\ (\ref{EQSm}) into Eq.\ (\ref{EQ2}).
Their presence is a consequence of $\Omega_c$ being a very complicated function of $\{\hat{e}_n\}$ beyond a Heisenberg picture, and they can be driving factors behind a free energy form which minimizes for first-order magnetic phase transformations~\cite{PhysRevB.99.144424}, as shown in Fig.\ \ref{FIG6}(a). High order terms in Eq.\ (\ref{EQSm}) have been conveniently arranged in order to produce multi-site local moment correlations in the free energy without repeating terms.

Our DFT-DLM codes are designed to provide the pairwise coefficients $\{\mathcal{S}_{aa,0}^{(2)},\mathcal{S}_{bb,0}^{(2)},\mathcal{S}_{ab,0}^{(2)}\}$ from a linear response of the paramagnetic state~\cite{0305-4608-15-6-018,PhysRevLett.82.5369}. They also give the site-dependent internal magnetic fields $\{\textbf{h}^\text{int}_n\}$ as functions of $\{\textbf{m}_n\}$~\cite{0305-4608-15-6-018},
\begin{equation}
\textbf{h}_{i}^\text{int}(\{\textbf{m}_n\}) \equiv-\nabla_{\textbf{m}_{i}}\big\langle\Omega(\{\hat{e}_n\})\big\rangle.
\label{EQh}
\end{equation}
A calculation of $\{\textbf{h}^\text{int}_n\}$ at many different states of ferromagnetic order, prescribed by $\{\textbf{m}=\frac{\textbf{m}_a+\textbf{m}_b}{2}\}$, enables a linear regression analysis of Eq.\ (\ref{EQh}), from which we obtain the dependence on $\{\textbf{m}_a,\textbf{m}_b\}$ given in Eq.\ (\ref{EQSm}) using Eq.\ (\ref{EQ2}). This enables higher order correlation coefficients to be obtained~\cite{PhysRevB.99.144424}.
The remaining entropy term in Eq.\ (\ref{EQ1}) to complete the free energy is calculated analytically by performing the following single-site integrals
\begin{equation}
\begin{split}
 & S_n(\lambda_n)=-k_\text{B}\int{\text{d}\hat{e}_n P_n(\hat{e}_n)\log P_n(\hat{e}_n)} \\
              &   =k_\text{B}\left[1+\log\left(4\pi\frac{\sinh\beta h^\text{int}_n}{\beta h^\text{int}_n}\right)-\beta h^\text{int}_n\coth \beta h^\text{int}_n\right].
\label{EQS}
\end{split}
\end{equation}

\begin{figure}[t]
\centering
\includegraphics[clip,scale=0.85]{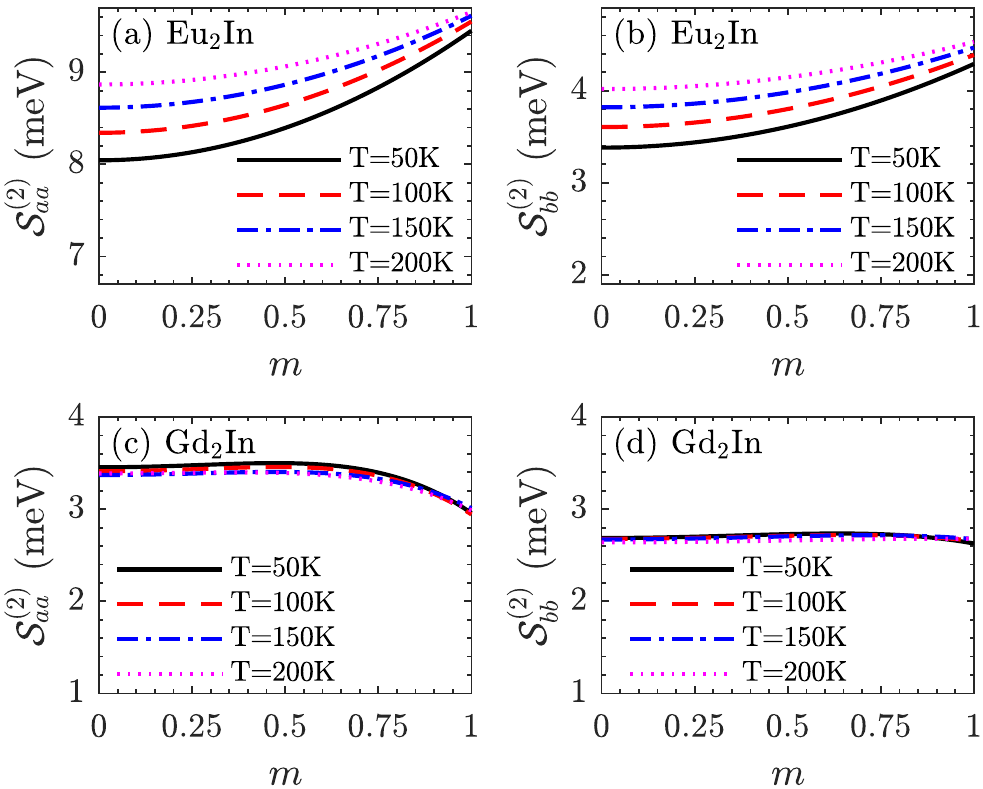}
\caption{$\mathcal{S}_{aa}^{(2)}$ amd $\mathcal{S}_{bb}^{(2)}$ as functions of $m=\left|\frac{\textbf{m}_a+\textbf{m}_b}{2}\right|$, as given in Eq.\ (\ref{EQSm}), obtained for (a,b) Eu$_2$In and (c,d) Gd$_2$In. Results are plotted for different temperatures, $T=50$K, 100K, 150K, and 200K, from which the $T$-dependence provided in table \ref{Tab1} has been calculated.}
\label{FIG7}
\end{figure}

The first-principles data of $\{\textbf{h}^\text{int}_n\}$ used to perform the analysis described above has been generated for values of $\{\textbf{m}_n\}$ ranging as $m_a=0\rightarrow0.9$ and $m_b=0\rightarrow0.9$ and several $m_a/m_b$ ratios. The nine higher than pairwise coefficients in Eq.\ (\ref{EQSm}), comprising coupling up to eighth order, were fitted using approximately 100 independent data points for each material. 
Their numerical error associated with the linear regression is in general 0.1meV or smaller.
This calculation was repeated for a series of temperatures ranging as $T=50,75,100,125,150,200$K to extract their temperature dependence.
The results obtained are shown in both Figs.\ \ref{FIG5}(a,b) and \ref{FIG7}, and table \ref{Tab1}, which show that $\mathcal{S}_{ab}^{(2)}$s in Fig.\ \ref{FIG5}(a,b) are the dominant correlations in both Eu$_2$In and Gd$_2$In.
We point out that for Gd$_2$In constant values are given in table \ref{Tab1} demonstrating a negligible temperature effect, as can also be seen in Fig.\ \ref{FIG7}(c,d).

\begin{table}
\begin{center}
\begin{tabular}{|c|c|c|}
\hline
Interaction & Eu$_2$In & Gd$_2$In \\[0.1cm]
\hline
$\mathcal{S}_{aa,0}^{(2)}$ & 7.78meV+0.00548$\frac{\text{meV}}{\text{K}}$ $\cdot T$(K) & 3.4meV \\[0.15cm]
$\mathcal{S}_{bb,0}^{(2)}$ & 3.18meV+0.00425$\frac{\text{meV}}{\text{K}}$ $\cdot T$(K) & 2.6meV \\[0.15cm]
$\mathcal{S}_{ab,0}^{(2)}$ & 21.1meV+0.00973$\frac{\text{meV}}{\text{K}}$ $\cdot T$(K) & 38meV \\[0.3cm]

$\mathcal{S}_{aa}^{(4)}$ & 1.62meV-0.00414$\frac{\text{meV}}{\text{K}}$ $\cdot T$(K) & -0.16meV \\[0.15cm]
$\mathcal{S}_{bb}^{(4)}$ & 1.05meV-0.00266$\frac{\text{meV}}{\text{K}}$ $\cdot T$(K) & 0.01meV \\[0.15cm]
$\mathcal{S}_{ab,1}^{(4)}$ & 9.01meV-0.0227$\frac{\text{meV}}{\text{K}}$ $\cdot T$(K) & 1.6meV \\[0.15cm]
$\mathcal{S}_{ab,2}^{(4)}$ & 7.59meV-0.0202$\frac{\text{meV}}{\text{K}}$ $\cdot T$(K) & -0.06meV \\[0.15cm]
$\mathcal{S}_{ab,3}^{(4)}$ & 8.98meV-0.0239$\frac{\text{meV}}{\text{K}}$ $\cdot T$(K) & -0.24meV \\[0.3cm]

$\mathcal{S}_{aa}^{(6)}$ & -24.8meV+0.0613$\frac{\text{meV}}{\text{K}}$ $\cdot T$(K) & -4.2meV \\[0.15cm]
$\mathcal{S}_{bb}^{(6)}$ & -3.42meV+0.00897$\frac{\text{meV}}{\text{K}}$ $\cdot T$(K) & 0.08meV \\[0.15cm]
$\mathcal{S}_{ab,1}^{(6)}$ & -4.49meV+0.0114$\frac{\text{meV}}{\text{K}}$ $\cdot T$(K) & 0.70meV \\[0.3cm]

$\mathcal{S}_{ab,1}^{(8)}$ & 12.48meV-0.0323$\frac{\text{meV}}{\text{K}}$ $\cdot T$(K) & 1.5meV \\[0.15cm]
\hline
\end{tabular}
\caption{Temperature-dependent local moment-local moment correlations numerically reproducing the \textit{ab-initio} data $\{\textbf{h}_n^\text{int}\}$ against extent of ferromagnetic order $\{\textbf{m}_a,\textbf{m}_b\}$ obtained from Eq.\ (\ref{EQh}) and using Eqs.\ (\ref{EQ2}) and (\ref{EQSm}). Results are shown for both orthorhombic Eu$_2$In and hexagonal Gd$_2$In. Since results for Gd$_2$In have a negligible dependence on $T$ we only show them at $T=150$K as a reference.}
\label{Tab1}
\end{center}
\end{table}

The minimization of the free energy $\mathcal{G}_1$ with respect to $\textbf{m}_a$ and $\textbf{m}_b$ at different temperatures provides the magnetization plots in Fig.\ \ref{FIG6}.
We found that the local order parameters that minimize $\mathcal{G}_1$ generally satisfy ratios equal to $r=m_a/m_b\approx 0.8$ and $r=m_a/m_b\approx 1$ for orthorhombic Eu$_2$In and hexagonal Gd$_2$In, respectively, at high temperature. The isothermal entropy changes for the evaluation of the magnetocaloric effect are calculated using Eq.\ (\ref{EQS}) and minimized curves of the Gibbs free energy at different values of an applied external magnetic field $\textbf{H}$~\cite{PhysRevB.99.144424,doi:10.1063/5.0003243}. The heat capacity is obtained using the same calculations and that $c=T\frac{\partial S}{\partial T}$.

To ensure that the exchange-correlation potential remedies the self-interaction of localized 4$f$-rare earth electrons, we introduce a local self-interaction correction (LSIC)~\cite{PhysRevB.71.205109,PhysRevMaterials.1.024411} in our treatment of Eu$_2$In and Gd$_2$In.
Owing to the spherical potential approximation of our multiple scattering formalism, the LSIC follows by correcting electronic states with spin and angular momentum quantum labels~\cite{PhysRevMaterials.1.024411}.
We model Eu$_2$In and Gd$_2$In in adherence to Hund's first rule, by applying the LSIC to half a shell of the 4$f$ electrons at rare-earth sites, which describes divalent and trivalent behaviors for Eu and Gd, respectively.
The corresponding DLM paramagnetic self-consistent potentials, whose local moment axis rotation was employed to generate the magnetic order dependent- ab initio data, were obtained using the Hutsepot code~\cite{D_ne_2009}.

In our calculations we employed a muffin-tin approximation. The maximum angular momentum value used in the expansions to solve the scattering single-site problems was $l_\text{max}=3$.
The lattice parameter values used were taken from experiment~\cite{Eu2In1,PALENZONA1968379} as $a=5.57$\AA, $b=10.30$\AA, and $c=7.44$\AA\, for Eu$_2$In, and $a=b=5.41$\AA, and $c=6.75$\AA\, for Gd$_2$In.

\section{Conclusions}
\label{conc}

In conclusion, we have found a purely electronic mechanism for an exceptional non-hysteretic first-order magnetic transition with a huge magnetocaloric response in Eu$_2$In. Our {\it ab initio} theory accurately describes the transition, heat capacity, and magnetocaloric trends. The magnetism of the itinerant valence electrons whereby a topological Fermi surface transition is triggered by spin polarization plays the central role. Near filling of bands with significant $d$-character and strong spin susceptibility enables the interactions between $f$-electron moments to deviate sharply from RKKY form and to strengthen as ferromagnetic order develops. For Eu$_2$In the favourable $E_\text{F}$ electronic structure is directly linked to strong $f$-electron correlations and Hund's first rule, which make Eu atoms divalent.  For the trivalent counterpart, Gd$_2$In, the effect is absent. We note that itinerant electron positive feedback also appears to drive the first-order magnetic transitions in other important magnetocaloric compounds such as La(Fe$_x$Si$_{1-x}$)$_{13}$~\cite{doi:10.1063/1.370471} and FeRh~\cite{PhysRevB.89.054427}. In these systems, however, the magnetic transitions have structural change repercussions which lead to detrimental thermal hysteresis~\cite{doi:10.1002/ente.201800264}. The present work consequently represents a theoretical benchmark for the search of hysteresis-free, discontinuous, magnetic phase transitions where itinerant electron response to and influence on  magnetic moment order replaces the need for magnetostructural coupling.

\begin{acknowledgments}

The present work forms part of the \href{https://warwick.ac.uk/fac/sci/physics/research/theory/research/electrstr/pretamag/}{PRETAMAG project}, funded by the UK Engineering and Physical Sciences Research Council, Grant No. EP/M028941/1.
E.\ M.-T.\ acknowledges funding from the DAAD short-term grant.
The research work at Ames is supported by the Critical Materials Institute, an Energy Innovation Hub funded by the U.S. Department of Energy, Office of Energy Efficiency and Renewable Energy, Advanced Manufacturing Office. The Ames Laboratory is operated for the U.S. Department of Energy by Iowa State University of Science and Technology under Contract No. DE-AC02-07CH11358.
This work madeuse of computational support by CoSeC, the ComputationalScience Centre for Research Communities, through CCP9.

\end{acknowledgments}

\bibliography{bibliography.bib}
\end{document}